\documentclass[preprint,showkeys,superscriptaddress,amsmath,amssymb]{revtex4-1}

\usepackage{xr}
\usepackage{marvosym}
\usepackage{amsmath}
\usepackage{times}
\usepackage{graphicx}
\usepackage{dcolumn}
\usepackage{color}
\usepackage[labelsep=space]{caption}
\usepackage{setspace}
\usepackage{amssymb}
\usepackage[version=3]{mhchem}
\usepackage{lineno}

\begin{document}

\renewcommand{\captionfont}{\small}
\newcommand{\NatureFormat}{%
		\renewcommand{\figurename}{\textbf{Figure}}
        \renewcommand{\thefigure}{\textbf{\arabic{figure}}}%
     }
\NatureFormat

\title{Compact Folded Metasurface Spectrometer}


\author{MohammadSadegh Faraji-Dana}
\email{These authors contributed equally to this work}
\affiliation{T. J. Watson Laboratory of Applied Physics and Kavli Nanoscience Institute, California Institute of Technology, 1200 E. California Blvd., Pasadena, CA 91125, USA}
\author{Ehsan Arbabi}
\email{These authors contributed equally to this work}
\affiliation{T. J. Watson Laboratory of Applied Physics and Kavli Nanoscience Institute, California Institute of Technology, 1200 E. California Blvd., Pasadena, CA 91125, USA}
\author{Amir Arbabi}
\affiliation{T. J. Watson Laboratory of Applied Physics and Kavli Nanoscience Institute, California Institute of Technology, 1200 E. California Blvd., Pasadena, CA 91125, USA}
\affiliation{Department of Electrical and Computer Engineering, University of Massachusetts Amherst, 151 Holdsworth Way, Amherst, MA 01003, USA}
\author{Seyedeh Mahsa Kamali}
\affiliation{T. J. Watson Laboratory of Applied Physics and Kavli Nanoscience Institute, California Institute of Technology, 1200 E. California Blvd., Pasadena, CA 91125, USA}
\author{Hyounghan Kwon}
\affiliation{T. J. Watson Laboratory of Applied Physics and Kavli Nanoscience Institute, California Institute of Technology, 1200 E. California Blvd., Pasadena, CA 91125, USA}
\author{Andrei Faraon}
\email{Corresponding author: A.F.: faraon@caltech.edu}
\affiliation{T. J. Watson Laboratory of Applied Physics and Kavli Nanoscience Institute, California Institute of Technology, 1200 E. California Blvd., Pasadena, CA 91125, USA}

\maketitle

\textbf{Recent advances in optical metasurfaces enable control of the wavefront, polarization and dispersion of optical waves beyond the capabilities of conventional diffractive optics. An optical design space that is poised to highly benefit from these developments is the “folded optics architecture” where light is confined between reflective surfaces and the wavefront is controlled at the reflective interfaces. In this manuscript we introduce the concept of folded metasurface optics by demonstrating a compact high resolution optical spectrometer made from a 1-mm-thick glass slab with a volume of 7 cubic millimeters. The spectrometer has a resolution of ~1.2 nm, resolving more than 80 spectral points in a 100-nm bandwidth centered around 810 nm. The device is composed of three different reflective dielectric metasurfaces, all fabricated in a single lithographic step on one side of a transparent optical substrate, which simultaneously acts as the propagation space for light. An image sensor, parallel to the spectrometer substrate, can be directly integrated on top of it to achieve a compact monolithic device including all the active and passive components. Multiple spectrometers, with similar or different characteristics and operation bandwidths may also be integrated on the same chip and fabricated in a batch process, significantly reducing their costs and increasing their functionalities and integration potential. In addition, the folded metasystems design can be applied to many optical systems, such as optical signal processors, interferometers, hyperspectral imagers and computational optical systems, significantly reducing their sizes and increasing their mechanical robustness and potential for integration.}

\begin{figure*}[htbp]
	\centering
	\fbox{\includegraphics[width=\linewidth]{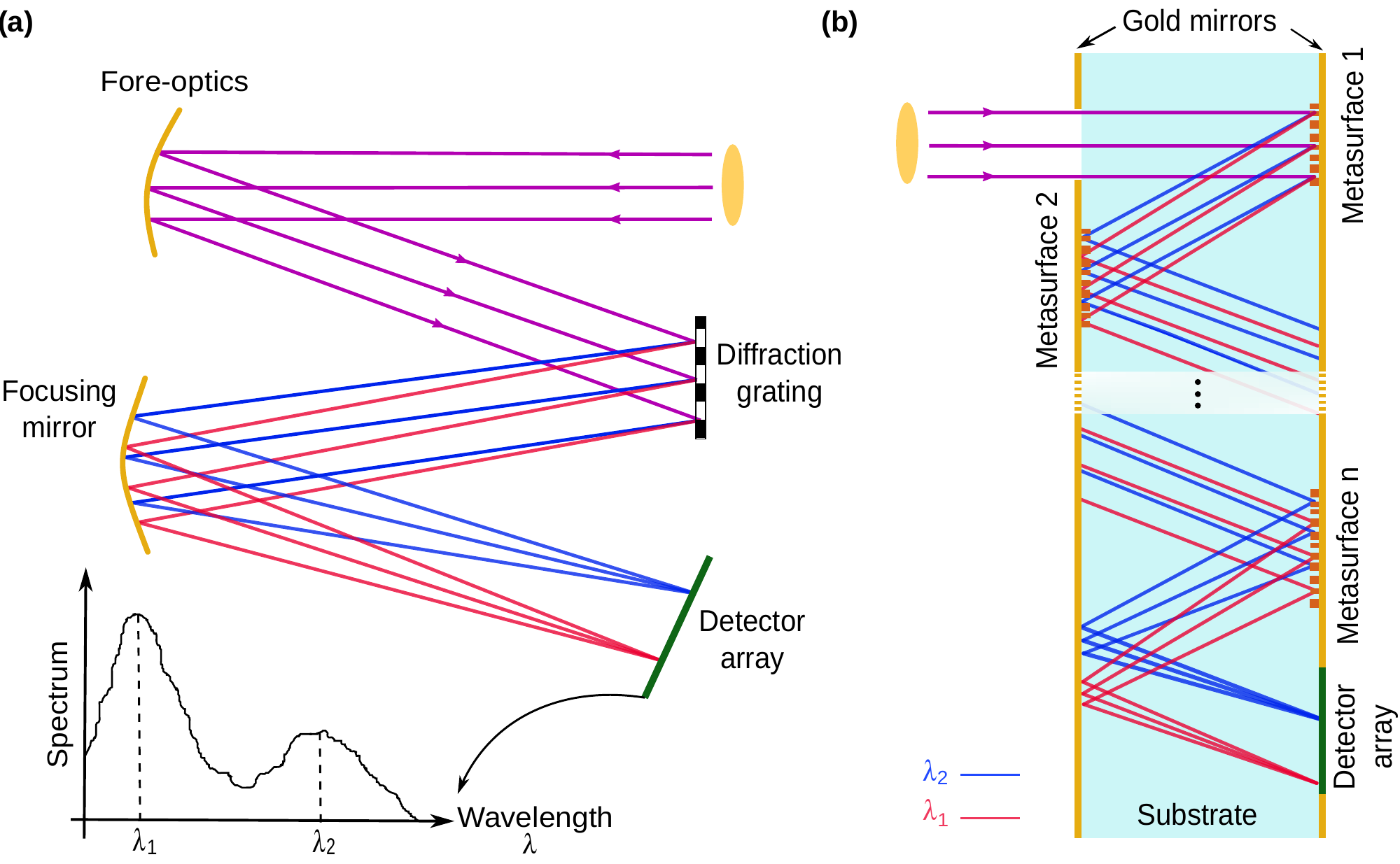}}
	\caption{\textbf{Schematics of a conventional and a folded metasurface spectrometer.} (\textbf{a}) Schematic illustration of a typical diffractive spectrometer. The main components are comprised of the fore-optics section, diffraction grating, focusing lenses and detector array. (\textbf{b}) The proposed scheme for a folded compact spectrometer. All the dispersive and focusing optics can be implemented as reflective metasurfaces on the two sides of a single transparent substrate. Mirrors on both sides confine and direct light to propagate inside the substrate, and the detector can be directly the output aperture of the device. If required, transmissive metasurfaces can also be added to the input and output apertures to perform optical functions. Although the schematic here includes metasurfaces on both sides to show the general case, the actual devices demonstrated here are designed to have metasurfaces only on one side to simplify their fabrication.}
	\label{fig:1_concept}
\end{figure*}

Optical spectrometry is a key technique in various areas of science and technology with a wide range of applications \cite{tkachenko2006optical,hollas2004modern}. This has resulted in a large demand for spectrometers and/or spectrum analyzers with different properties (e.g., operation bandwidth, resolution, size, etc.) required for different applications \cite{nevejans2006compact, thomas1966optical,azana2000real}. Conventional optical spectrometers are composed of a dispersive element, such as a prism or a diffraction gating, that deflects different wavelengths of light by different angles, followed by focusing elements that focus light incoming from different angles to different points (or lines). As schematically shown in Fig. ~\ref{fig:1_concept}a , the intensity at different wavelengths can then be measured using an array of detectors. Diffraction gratings have typically larger dispersive powers than transparent materials, and therefore diffractive spectrometers generally have better resolutions \cite{tkachenko2006optical}. The combination of several free space optical elements (the grating, focusing mirrors, etc.) and the free space propagation volume result in bulky spectrometers. 

In recent years, there has been an increased interest in high-performance compact spectrometers that can be easily integrated into consumer electronics for various medical and technological applications such as medical diagnosis, material characterization, quality control, etc. \cite{ferrari2012brief,momeni2009integrated} . As a result, various schemes and structures have been investigated for realization of such spectrometers \cite{wang2007concept,pervez2010photonic,xia2011high,redding2013compact,nitkowski2008cavity,momeni2009integrated,gan2012high,bockstaele2011compact,shibayama2013spectroscope,grabarnik2008high}. One class of miniaturized spectrometers integrate a series of band-pass filters with different center wavelengths on an array of photodetectors~\cite{wang2007concept,horie2016wide}. Although these devices are compact and compatible with standard microfabrication techniques, they have resolutions limited by achievable filter quality factors, and low sensitivities caused by the filtering operation that rejects a large portion of the input power. Spectrometers based on planar on-chip integrated photonics provide another solution with high spectral resolution \cite{xia2011high,redding2013compact,nitkowski2008cavity,momeni2009integrated,gan2012high,pervez2010photonic}. However, the loss associated with on-chip coupling of the input light and the reduced throughput because of the single-mode operation \cite{madi2018lippmann} are still major challenges for widespread adoption in many applications.
Another type of compact spectrometers are conceptually similar to the conventional table-top spectrometers, however, they use micro-optical elements to reduce size and mass~\cite{bockstaele2011compact,shibayama2013spectroscope}. Due to the inferior quality and limited control achievable by micro-optical elements as well as the shorter optical path lengths, these devices usually have lower spectral resolutions. Higher resolution has been achieved by using aberration-correcting planar gratings ~\cite{grabarnik2008high}, however an external spherical mirror makes the device bulky.

Dielectric metasurfaces, a new category of diffractive optical elements with enhanced functionalities, have attracted a great deal of interest in recent years ~\cite{lalanne2017metalenses,kruk2017functional,hsiao2017fundamentals,lin2014dielectric}. Overcoming many of the material and fundamental limitations of plasmonic metasurfaces~\cite{arbabi2017fundamental}, dielectric metasurfaces have proven capable of implementing several conventional ~\cite{Arbabi2016NatCommun,arbabi2015subwavelength, kamali2016highly,lalanne2017metalenses,fattal2010flat,Chong2015NanoLett,zhou2017efficient,zhan2016low,yang2014dielectric,chu2016active,yu2014flat,paniagua2018metalens} and new optical devices~\cite{Arbabi2015NatNano,backlund2016removing,kamali2016decoupling,kamali2017angle,yang2015nonlinear} with high efficiencies. They enable control of phase with subwavelength resolution and high gradients and simultaneous control of phase and polarization~\cite{Arbabi2015NatNano}. A key feature of metasurfaces is their compatibility with micro and nano-fabrication techniques, which allows for integration of multiple metasurfaces for realizing complex optical metasystems~\cite{Arbabi2016NatCommun,arbabi2017planar}. Such metasystems allow for significantly improving optical properties of metasurfaces through aberration correction (such as lenses with diffraction limited operation over wide field of view ~\cite{Arbabi2016NatCommun}), or functionalities fundamentally unachievable with local single-layer metasurfaces such as retroreflection ~\cite{arbabi2017planar}.

Taking a different approach to device integration, here we introduce folded optical metasystems where multiple metasurfaces are integrated on a single substrate that is also playing the role of propagation space for light [Fig. \ref{fig:1_concept}b]. Using this platform, we experimentally demonstrate a compact folded optics device for spectroscopy with a 1-mm thickness ($\sim$7-mm$^3$ volume) that provides a $\sim$1.2-nm resolution over a 100-nm bandwidth (more than 80 points over a $\sim$12$\%$ bandwidth) in the near infrared. As schematically shown in Fig. \ref{fig:1_concept}b, multiple reflective metasurfaces can be designed and fabricated on the same transparent substrate to disperse and focus light to different points on a plane parallel to the substrate. To the best of our knowledge, this is the first demonstration of an optical metasystem comprising more than two metasurfaces that implements a sophisticated optical functionality like spectrometry. Furthermore, the presented configuration can allow for the integration of the detector array on top of the folded spectrometer, resulting in a compact monolithic device. We should note here that it was recently demonstrated that an off-axis metasurface lens (i.e., a lens with an integrated blazed grating phase profile~\cite{khorasaninejad2016super,zhu2017ultra}) can disperse and focus different wavelengths to different points. However, there are fundamental and practical limitations for such elements that significantly limits their application as a spectrometer (which is the reason why other types of diffractive optical elements, such as holographic optical elements and kinoforms, that can essentially perform the same function have not been used for this application before). Fundamentally, the chromatic dispersion~\cite{Born1999,miyamoto1961design,Faklis1995ApplOpt,Arbabi2016Optica,arbabi2017controlling} and angular response correlation of diffractive optical elements and metasurfaces~\cite{jang2018wavefront,kamali2017angle} limit the bandwidth and angular dispersion range where the device can provide tight aberration-free focusing. This in turn limits the achievable resolution and bandwidth of the device. Moreover, the chromatic dispersion results in a focal plane almost perpendicular to the metasurface, which will then require the photodetector array to be placed almost normal to the metasurface plane \cite{zhang2016g,khorasaninejad2016super,zhu2017ultra}. In addition to the distance for the propagation of dispersed light, this normal placement undermines the compactness of the device.

\begin{figure*}[htbp]
	\centering
	\fbox{\includegraphics[width=\linewidth]{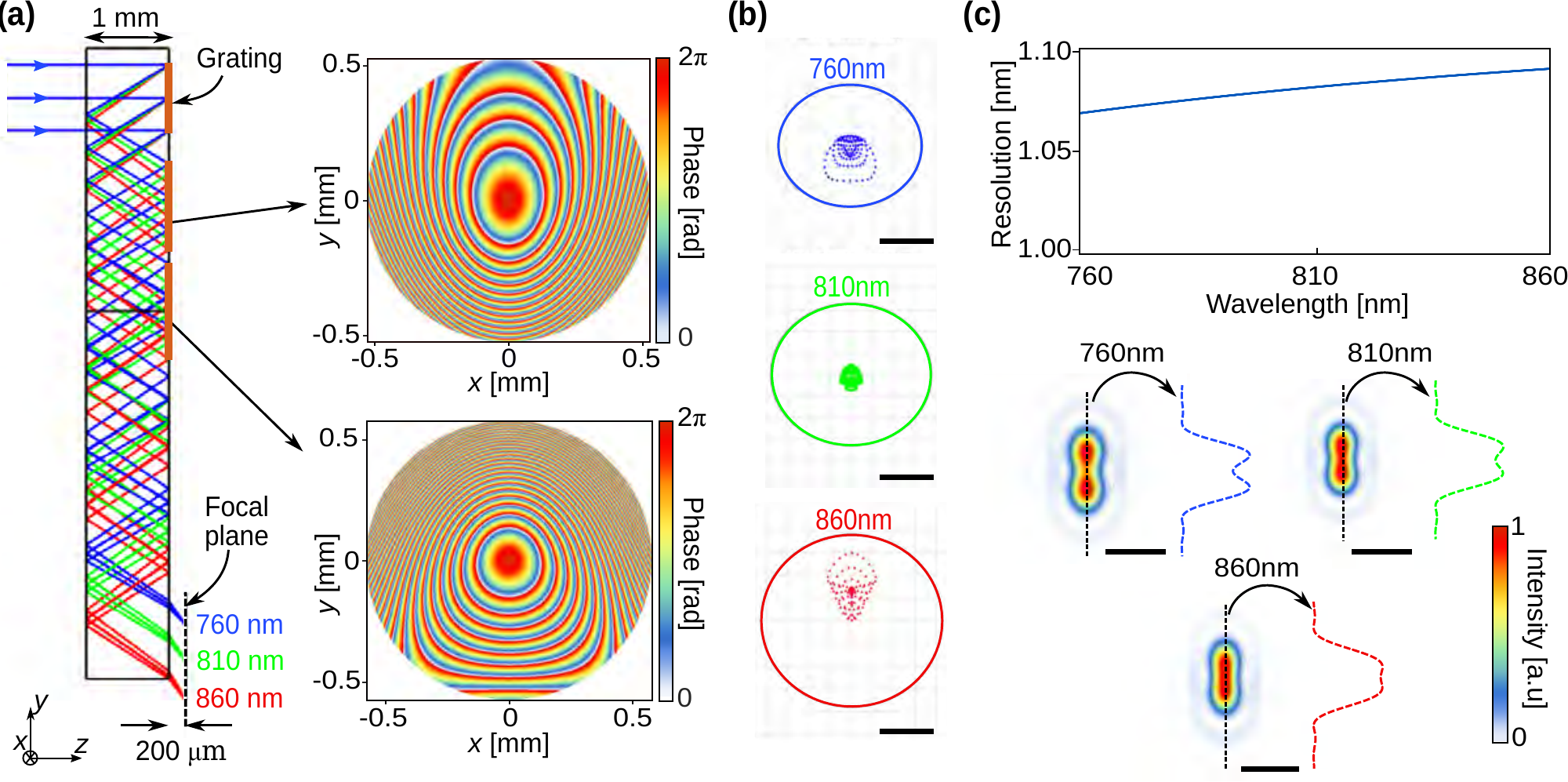}} 
	\caption{\textbf{Ray-optics design and simulation results of the folded spectrometer.} (\textbf{a}) Ray tracing simulation results of the folded spectrometer, shown at three wavelengths in the center and two ends of the band. The system consists of a blazed grating that disperses light to different angles, followed by two metasurfaces optimized to focus light for various angles (corresponding to different input wavelengths). The grating has a period of 1~$\mu$m, and the optimized phase profiles for the two metasurfaces are shown on the right. (\textbf{b}) Simulated spot diagrams for three wavelengths: center and the two ends of the band. The scale bars are 5~$\mu$m. (\textbf{c}) Spectral resolution of the spectrometer, which is calculated from simulated Airy disk radii and the lateral displacement of the focus with wavelength, is shown on the top. Bottom: simulated intensity distribution for two wavelengths separated by 1.1 nm around three different center wavelengths of 760 nm, 810 nm, and 860 nm. The intensity distributions show that wavelengths separated by 1.1 nm are theoretically resolvable. The scale bars are 20~$\mu$m.}
	\label{fig:2_zemax}
\end{figure*}

Figure \ref{fig:2_zemax}a shows the ray tracing simulations of the designed spectrometer. The device consists of three metasurfaces, all patterned on one side of a 1-mm-thick fused silica substrate. The first metasurface is a periodic blazed grating with a period of 1~$\mu$m that disperses different wavelengths of a collimated input light to different angles, centered around 33.9$^{\circ}$ at 810 nm. The second and third metasurfaces focus light coming from different angles (corresponding to various input wavelengths) to different points on the focal plane. We have recently demonstrated a metasurface doublet capable of correcting monochromatic aberrations to achieve near-diffraction-limited focusing over a wide field of view~\cite{Arbabi2016NatCommun}. The second and third metasurfaces here essentially work similar to the mentioned doublet, with the difference of working off axis and being designed in a folded configuration, such that the focal plane for our desired bandwidth is parallel to the substrate. To simplify the device characterization, the focal plane was designed to be located 200~$\mu$m outside the substrate. The asymmetric design of the focusing metasurfaces in an off-axis doublet configuration, allows for the focal plane to be parallel to the substrate. This makes the integration of the spectrometer and the detector array much simpler, results in a more compact and mechanically robust device, and allows for direct integration into consumer electronic products like smartphones. The optimized phase profiles for the two surfaces are shown in Fig. \ref{fig:2_zemax}a, right (see Supplementary Table I for the analytical expression of the phase). Simulated spot diagrams of the spectrometer are plotted in Fig. \ref{fig:2_zemax}b for three wavelengths at the center and the two ends of the bandwidth showing negligible geometric aberrations. The spot diagrams are plotted only at three wavelengths, but the small effect of optical aberrations was confirmed for all wavelengths in the 760 nm-860 nm bandwidth. As a result, the spectral resolution of the device can be calculated using the diffraction limited Airy radius and the lateral displacement of the focus by changing the wavelength. The calculated resolution is plotted in Fig. \ref{fig:2_zemax}c, showing a theoretical value of better than 1.1 nm across the band. Point spread functions (PSFs) calculated for input beams containing two wavelengths 1.1 nm apart, and centered at 760 nm, 810 nm, and 860 nm are plotted in the same panel, showing two resolvable peaks.

\begin{figure*}[htbp]
	\centering
	\fbox{\includegraphics[width=\linewidth]{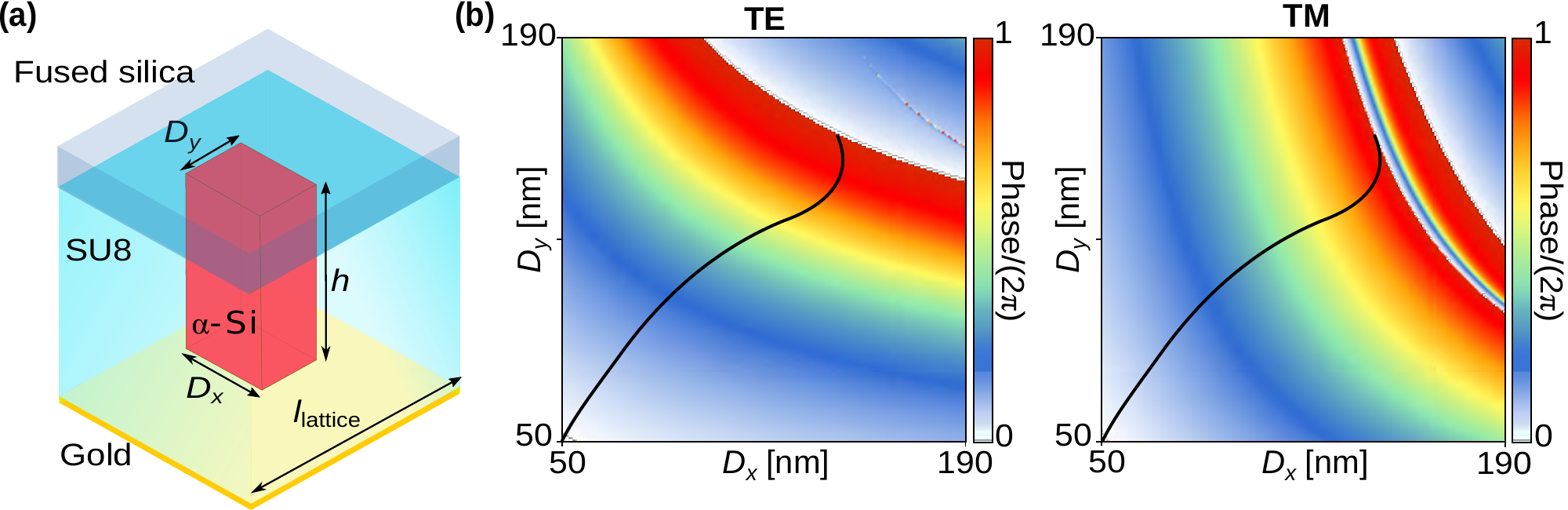}}
	\caption{\textbf{Metasurface structure and design graphs.} (\textbf{a}) Schematics of the reflective rectangular meta-atom. The meta-atom consists of $\alpha$-Si nano-posts on a fused silica substrate, capped by a layer of SU-8 polymer and backed by a gold mirror. The nano-post is 395 nm tall and the lattice constant is 250~nm for the blazed grating and 246~nm for the focusing metasurfaces. Schematics of the simulated structure and conditions are shown on the bottom. (\textbf{b}) Simulated reflection phase plotted for TE and TM polarizations. The black curve highlights the path through the $D_\mathrm{x}$-$D_\mathrm{y}$ plane that results in equal phases for the two polarizations. Nano-posts on this path were used to design the two focusing metasurface elements to make them insensitive to the input polarization.}
	\label{fig:3_PhaseProfile}
\end{figure*}

To implement the reflective metasurfaces, we used a structure similar to the reflective elements in~\cite{arbabi2017planar}. Each of the meta-atoms, shown schematically in Fig.~\ref{fig:3_PhaseProfile}a, consists of an $\alpha$-Si nano-post with a rectangular cross section, capped by a 2-$\mu$m-thick SU-8 layer and backed by a gold mirror. The post height and lattice constant were chosen to be 395 nm and 246 nm, respectively, to achieve full 2$\pi$ phase coverage while minimizing variation of the reflection phase derivative across the band (Supplementary Fig. \textbf{S1}). Minimizing the phase derivative variation will mitigate the reduction of device efficiency over the bandwidth of interest~\cite{arbabi2017controlling} by decreasing the wavelength dependence of the phase profiles (Supplementary Fig. \textbf{S2}). In addition, since the two focusing metasurfaces are working under an oblique illumination ($\theta\sim$33.9$^\circ$), the nano-posts were chosen to have a rectangular cross-section to minimize the difference in reflection amplitude and phase for the transverse electric (TE) and transverse magnetic (TM) polarizations (for the oblique incident angle of $33.9^\circ$ at 810 nm). Reflection coefficients are found through simulating a uniform array of nano-posts under oblique illumination ($\theta\sim$33.9$^\circ$) with TE and TM polarized light [Fig. \ref{fig:3_PhaseProfile}a, right]. The simulated reflection phase as a function of the nano-posts side lengths are shown in Fig. \ref{fig:3_PhaseProfile}b. The black triangles highlight the path through the $D_\mathrm{x}$-$D_\mathrm{y}$ plane along which the reflection phase for the TE and TM polarizations is almost equal. In addition, as shown in Supplementary Fig. \textbf{S2} having almost the same reflection phases for the TE and TM polarizations holds true for the whole desired 760nm-860 nm bandwidth. The nanopost dimensions calculated from this path were used to implement the two focusing metasurfaces. 

The blazed grating has a periodic phase profile (with a period of 1 $\mu$m) that deflects normally incident light to a large angle inside the substrate. With a proper choice of the lattice constant (250~nm, in our case), its structure can also be periodic. This different structure and operation require a different design approach. The periodicity of the grating allows for its efficient full-wave simulation which can be used to optimize its operation over the bandwidth of interest. A starting point for the optimization was chosen using the recently developed high-NA lens design method~\cite{arbabi2017increasing}, and the structure was then optimized using the particle swarm optimization algorithm to simultaneously maximize deflection efficiency at a few wavelengths in the band for both polarizations (see Supplementary Section \textbf{S1} and Fig. \textbf{S3} for details).

\begin{figure*}
	\centering 
	\fbox{\includegraphics[width=\linewidth]{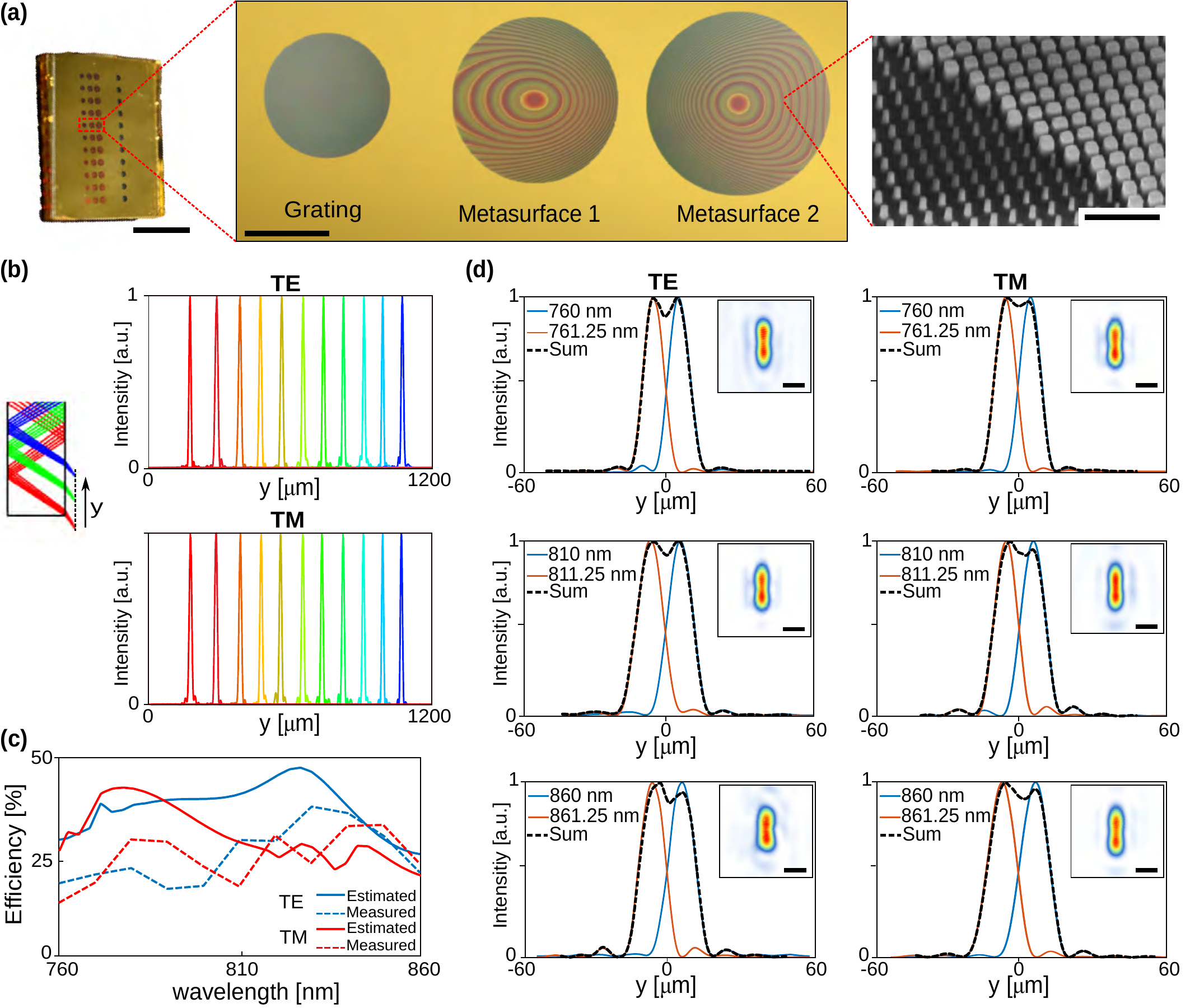}}
	\caption{\textbf{Experimental characterization results.} (\textbf{a}) An optical microscope image of the fabricated device and metasurfaces before deposition of the second gold layer. Inset (right) shows a scanning electron micrograph of a portion of one of the two focusing metasurfaces (scale bars from left to right: 10 mm, 500~$\mu$m, and 1~$\mu$m). (\textbf{b}) One dimensional focal spot profiles measured for several wavelengths in the bandwidth along the $y$-direction (as indicated in the inset) for TE (top) and TM (bottom) polarizations. The wavelengths start at 760 nm (the right most profile, blue curve) and increase at 10-nm steps up to 860 nm (the left most profile, red curve). (\textbf{c}) Calculated and measured absolute focusing efficiencies of the spectrometer for TE and TM polarizations. Both polarizations have average measured efficiencies of $\sim$25$\%$. (\textbf{d}) Measured intensity distributions for two input wavelengths that are 1.25 nm apart. The measurements were carried out at the center and and the two ends of the wavelength range for both polarizations. The insets show the corresponding 2-dimensional intensity profiles, demonstrating two resolvable peaks (scale bars: 10~$\mu$m).} 
	\label{fig:4_DevResp}
\end{figure*}

The device was fabricated using conventional micro- and nano-fabrication techniques. First, a 395-nm-thick layer of $\alpha$-Si was deposited on a 1-mm-thick fused silica substrate. All metasurfaces were then patterned using electron beam lithography in a single step, followed by a pattern inversion through the lift-off and dry etching processes. The metasurfaces were capped by a $\sim$2-$\mu$m-thick SU-8 layer, and a 100-nm-thick gold layer was deposited as the reflector. A second reflective gold layer was deposited on the second side of the substrate. Both the input and output apertures (with diameters of 790 $\mu$m and 978 $\mu$m, respectively) were defined using photolithography and lift-off. An optical microscope image of the three metasurfaces, along with a scanning electron micrograph of a part of the fabricated device are shown in Fig. \ref{fig:4_DevResp}a.

To experimentally characterize the spectrometer, a normally incident collimated beam from a tunable continuous wave laser was shinned on the input aperture of the device. A custom-built microscope was used to image the focal plane of the spectrometer, $\sim$ 200$\mu$m outside its output aperture (see Supplementary Section S1 and Fig. \textbf{S5} for details of the measurement setup). The input wavelength was tuned from 760 nm to 860 nm in steps of 10 nm, and the resulting intensity distributions were imaged using the microscope. The resulting one-dimensional intensity profiles are plotted in Fig.~\ref{fig:4_DevResp}b for TE (left) and TM (right) polarizations. The intensity profiles were measured over the whole 1.2-mm length of the y-direction in the focal plane (as shown in Fig.~\ref{fig:4_DevResp}b, inset) at each wavelength. The background intensity is beyond visibility in the linear scale profiles plotted here for all wavelengths (see Supplementary Figs. S6 and S7 for two-dimensional and logarithmic-scale plots of the intensity distribution, respectively). Figure~\ref{fig:4_DevResp}d shows the measured intensity profiles for three sets of close wavelengths, separated by 1.25 nm. The insets show the corresponding two-dimensional intensity distribution profiles. For all three wavelengths, and for both polarizations the two peaks are resolvable. The experimentally obtained spectral resolution is plotted in Supplementary Fig. \textbf{S8} versus the wavelength. The average resolution for both polarizations is $\sim$1.2~nm, which is slightly worse than the theoretically predicted value ($\sim$1.1~nm). We attribute the difference mostly to practical imperfections such as the substrate having an actual thickness different from the design value and thickness variation. In addition the metasurface phases are slightly different from the designed profiles due to fabrication imperfections. The angular sensitivity/tolerance of the device was also measured with respect to polar and azimuthal angle deviations from 0 incidence angle, in the $x$-$z$ and $y$-$z$ planes, using the setup shown in Supplementary Fig. \textbf{S9}c). In the $y$-$z$ plane the maximum tilt angle to maintain the same 1.25 nm resolution is $\pm$0.15$^{\circ}$, while in the $x$-$z$ plane the device has a $\pm$1$^{\circ}$ degree acceptance angle. The measurement results in Fig. \textbf{S9} match well with the predictions from ray-tracing simulations.

The measured and calculated focusing efficiencies are plotted in Fig. ~\ref{fig:4_DevResp}c. The focusing efficiency, defined as the power passing through a $\sim$30-$\mu m$ diameter pinhole around the focus divided by the total power hitting the input aperture, was measured using the setup shown in Supplementary Fig. \textbf{S5}. For both polarizations, the average measured efficiency is about 25$\%$. As seen from the measured efficiency curves, the optimization of the blazed grating efficiency versus wavelength and the choice of the design parameters to minimize variations in the phase-dispersion for the doublet metasurface lens, have resulted in a smooth measured efficiency. An estimate for the expected efficiency (shown as simulated efficiency in Fig.~\ref{fig:4_DevResp}c ) is calculated by multiplying the deflection efficiency of the grating, the efficiency of seven reflections off the gold mirrors, the input and output aperture transmission efficiencies, and the average reflectivities of the uniform nano-post arrays (as an estimate for the two focusing metasurface efficiencies). It is worth noting that considering only the reflection losses at the interfaces (nine reflective ones, and two transmissive ones) reduces the efficiency to about 48$\%$, showing a close to 50$\%$ efficiency for the three metasurfaces combined. We attribute the remaining difference between the measured and estimated values to fabrication imperfections (e.g., higher loss for the actual gold mirrors, and imperfect fabrication of the metasurfaces), the lower efficiency of the metasurfaces compared to the average reflectivity of uniform arrays, and to the minor difference from the designed value of the metasurface phase profiles at wavelengths other than the center frequency.

\begin{figure*}
	\centering
	\fbox{\includegraphics[width=\linewidth]{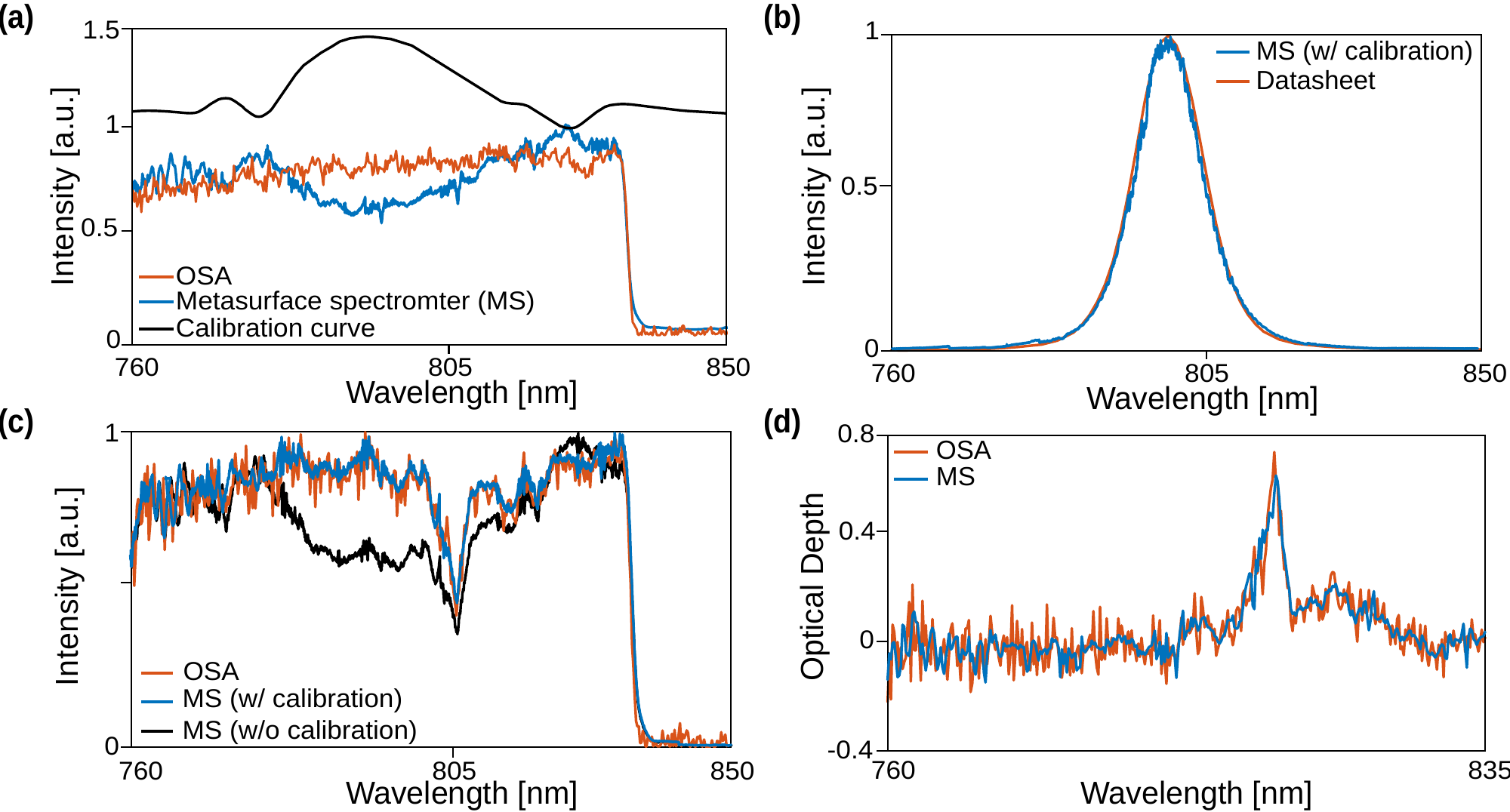}}
	\caption{\textbf{Sample spectrum measurement results.} (\textbf{a})  Spectrum of a wideband source (a super-continuum laser with an 840-nm short-pass filter) measured by a commercial OSA and the metasurface spectrometer (MS). This measurement was used to extract the calibration curve. (\textbf{b}) Spectrum of a 10-nm band-pass filter centered at 800 nm measured by the MS, compared to the spectrum acquired from the filter datasheet. (\textbf{c}) Transmission spectrum of a $\text{Nd}:\text{YVO}_4$ crystal sample measured both with a commercial OSA and the MS.  (\textbf{d}) The optical depth of the sample extracted from the spectrum measurements both with the OSA and the MS.}
	\label{fig:5_MeasSpectrum}
\end{figure*}

To demonstrate that the metasurface spectrometer actually has the ability to measure dense optical spectra, we use it to measure the transmission spectra of two different samples. First, we measured the spectrum of a wideband source (a super-continuum laser source, filtered with an 840-nm short-pass filter), both with the metasurface spectrometer (MS) and a commercial optical spectrum analyzer (OSA). By dividing the spectra measured by the two devices, we extract the required calibration curve that accounts for the variation of the metasurface spectrometer as well as the non-uniformities in the responsivity of the optical setup used to image the focal plane (i.e., the objective lens and the camera, as well as the optical fiber used to couple the signal to the OSA). The measured spectra and the extracted calibration curve are plotted in Fig.~\ref{fig:5_MeasSpectrum}a. Next, we used this calibration curve to measure the transmission spectrum of a band-pass filter with a nominal 10-nm full width at half maximum bandwidth and centered at 800~nm. The measured spectrum along with the transmission spectrum obtained from the filter datasheet are plotted in Fig. ~\ref{fig:5_MeasSpectrum}b, showing a good agreement. Finally, we used the metasurface spectrometer to measure the optical depth of a Nd:YVO$_4$ crystal sample. The spectrum measured with the metasurface spectrometer (after calibration) is compared with the transmission spectrum of the same sample measured with the OSA in Fig.~\ref{fig:5_MeasSpectrum}c. Dividing the spectrum without and with the sample, we have extracted the optical depth of the sample which is plotted in Fig.~\ref{fig:5_MeasSpectrum}d. A good agreement is observed between the two measurement results. It is worth noting that the Nd:YVO$_4$ crystal sample was cut though the z-plane, resulting in an equal absorption spectrum for the two polarizations. Therefore, we can assume that all spectral measurements were done with the same state of input polarization. This justifies the use of only one calibration curve for all the measurements.

The measured efficiency of the spectrometer demonstrated here is about 25$\%$. This value can be significantly increased to about 70$\%$ by using mirrors with higher reflectivity (e.g., DBRs or high contrast grating mirrors~\cite{bekele2015polarization,chang2012high}), and anti-reflection coatings on the input and output apertures. In addition, more advanced optimization techniques~\cite{sell2017large} could be exploited to optimize the diffraction grating to achieve high efficiency and polarization insensitivity. Implementing these changes and optimizing the fabrication process, one can expect to achieve efficiencies exceeding 70$\%$ for the spectrometer.

The metasurface spectrometers are fabricated in a batch process, and therefore many of them can be fabricated on the same chip, even covering multiple operation bandwidths. This can drastically reduce the price of these devices, allowing for their integration into various types of systems for different applications. In addition, the demonstrated structure is compatible with many of the techniques developed for the design of multi-wavelength metasurfaces~\cite{Arbabi2016SciRep,Arbabi2016Optica}, and therefore one might be able to combine different optical bandwidths into the same device (e.g., using a grating that deflects to the right at one bandwidth, and to the left at the other), resulting in compact devices with enhanced functionalities.

The optical throughput (etendue) is a fundamental property of any optical system, setting an upper limit on the ability of the system to accept light from spatially incoherent sources. It can be estimated as the product of the physical aperture size and the acceptance solid angle of the system. Furthermore, the total etendue of a system is limited by the element with the lowest etendue. To calculate the throughput of the metasurface spectrometer, we have performed simulations and measurements to characterize its acceptance angle. According to the measurement results in Supplementary Fig. \textbf{S9} the acceptance angle of the system is about 2 degrees in the horizontal direction, and 0.3 degrees in the vertical direction. Given this and the input aperture dimensions, the optical throughput of our device is calculated to be  $\sim$90 $Sr(\mu m)^2$. For comparison, the etendue of optical systems operating around 1 $\mu$m that utilize single-mode input channels (i.e., most optical spectrometers based on integrated optics platforms) is around $\sim$1 $Sr (\mu m)^2$. Furthermore, the demonstrated spectrometer is optimized for maximum sensitivity and not throughput. To show that the achieved throughput here does not denote an upper limit for the etendue of a folded metasurface spectrometer with similar characteristics (i.e., resolution, bandwidth, etc.), we have designed a second device with a throughput of $\sim$13000 $Sr (\mu m)^2$ (see Supplementary section \textbf{S2} and Fig. \textbf{S10} for design details and simulation results). Table \ref{tab:throughput} provides a comparison of the optical throughput of several compact spectrometers from recent literature. According to this table, the spectrometers designed using the folded metasurface platform can collect 2 to 4 orders of magnitude more light compared to on-chip spectrometers that are based on single/few-mode input waveguides, resulting in a much higher sensitivity.

\begin{table*}[htbp]
	\centering
	\caption{\bf Comparison of different spectrometers in terms of throughput (Etendue) and their dimensions}
	
	\scalebox{0.98}{
		\begin{tabular}{|c|c|c|}
			\hline
			Spectrometer & Etendue [$Sr(\mu m)^2$] & Size (dimensions) \\
			\hline
			\cite{redding2013compact}   & $<$0.5   & $50\mu m \times 100\mu m \times$ thickness  \\
			\cite{gatkine2017arrayed} & $\sim$ 0.8 & $16 mm \times 7 mm \times 15\mu m$\\
			This work & $\sim$ 90  & $1mm \times 1 mm \times 7 mm$\\
			\cite{shibayama2013spectroscope} & 8250 & $20.1 mm \times 12.5 mm \times 10.1 mm$\\
			Extension in Fig. \textbf{S10} & $\sim$13000 & $2mm \times 2.5 mm \times 8 mm$ \\
			\hline
	\end{tabular}}
	\label{tab:throughput}
\end{table*}

The development of thin and compact optical elements and systems has been a key promise of optical metasurfaces. Although many optical devices have been developed in thin and compact form factors using metasurfaces, significantly reducing the volume of optical systems using metasurfaces has not been previously demonstrated due to the requirement of the free-space propagation in many systems (e.g., imaging systems, spectrometers, etc.). The folded metasystem configuration introduced here can significantly reduce the size of many of these optical systems using the substrate as the propagation space for light. Based on this configuration, we demonstrated a 1-mm-thick spectrometer with a 7-mm$^3$ volume, reduced by a factor of ten compared to the same system implemented in an unfolded scheme (twenty times reduction, if the same system was designed in air). The spectrometer has a resolution of $\sim$1.2 nm over a 100-nm bandwidth ($>$12$\%$) in the near infrared. Using this design, multiple spectrometers can be fabricated on the same chip and in the same process, significantly reducing the costs and enabling integration of spectrometers covering multiple optical bands into consumer electronics. Moreover, by improving the angular response of the current device one can design a compact hyperspectral imager capable of simultaneous one-dimensional imaging and spectroscopy. In a broader sense, we expect that the proposed platform will also be used for on-chip interferometers, imaging systems, and other devices performing complex transformations of the field.

\clearpage
\section*{Acknowledgement}
This work was supported by Samsung Electronics. M.F. was partly supported by The Natural Sciences and Engineering Research Council of Canada (NSERC). S.M.K.is supported as part of the Department of Energy (DOE) ``Light-Material Interactions in Energy Conversion‚" Energy Frontier Research Center under grant no. DE-SC0001293. The device nano-fabrication was performed at the Kavli Nanoscience Institute at Caltech. Authors would like to thank Dr. Tian Zhong for providing the Nd:YVO$_4$ sample, and Dr. Seunghoon Han and Dr. Duhyun Lee for useful discussions.

\clearpage

\bibliographystyle{naturemag}
\bibliography{References}

\clearpage

\section*{Figures}

\end{document}


{\Large\textbf{Compact Folded Metasurface Spectrometer: supplementary material}}\\

\author{MohammadSadegh Faraji-Dana}
\author{Ehsan Arbabi}
\author{Amir Arbabi}
\author{Seyedeh Mahsa Kamali}
\author{Hyounghan Kwon}
\author{Andrei Faraon}
\maketitle
\renewcommand{\figurename}{Figure}
\renewcommand{\thefigure}{\textbf{S\arabic{figure}}}

\begin{figure*}[htbp]
	\centering
	\fbox{\includegraphics[width=\linewidth]{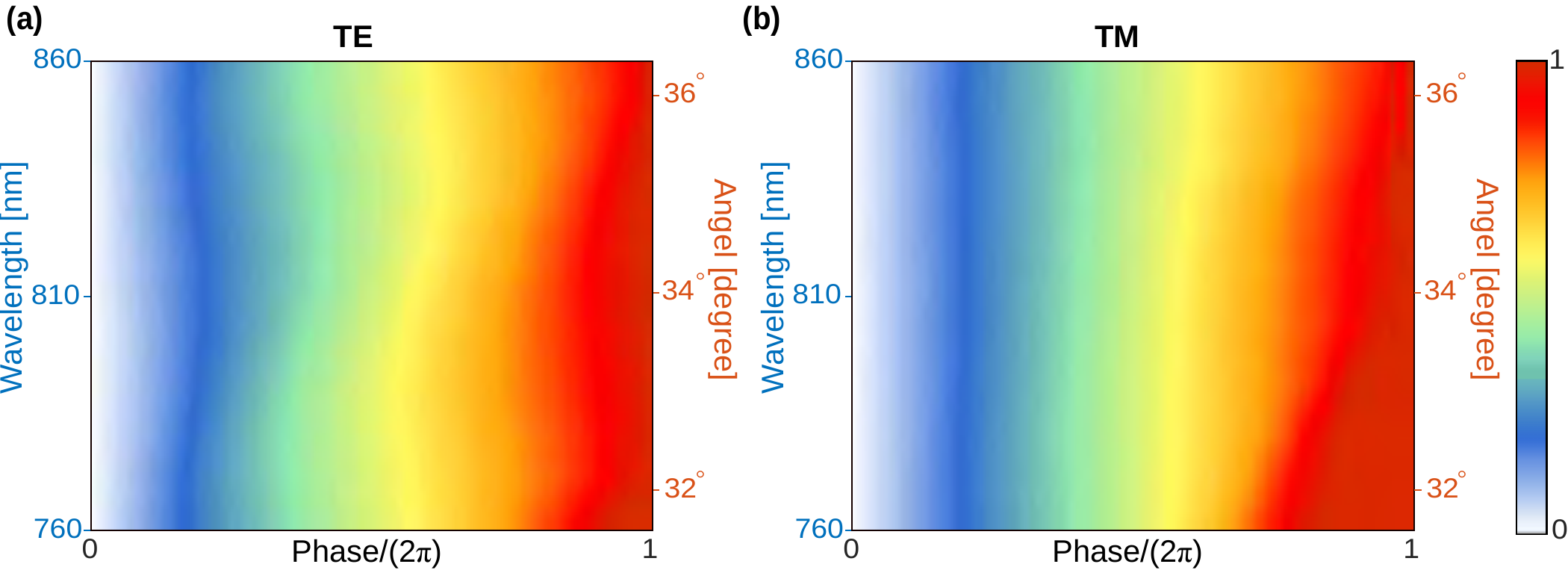}}
	\caption{\textbf{Phase variation vs wavelengths and input incident angles.} (\textbf{a}) Phase achieved by nano posts in Fig. \textbf{3} versus wavelengths and incident angles created by the diffraction grating for TE polarization (\textbf{b}) The same graph plotted for TM polarization.}
	\label{fig:S2Pre_PhaseLambdaAngle}
\end{figure*}

\begin{figure*}[htbp]
	\centering
	\fbox{\includegraphics[width=\linewidth]{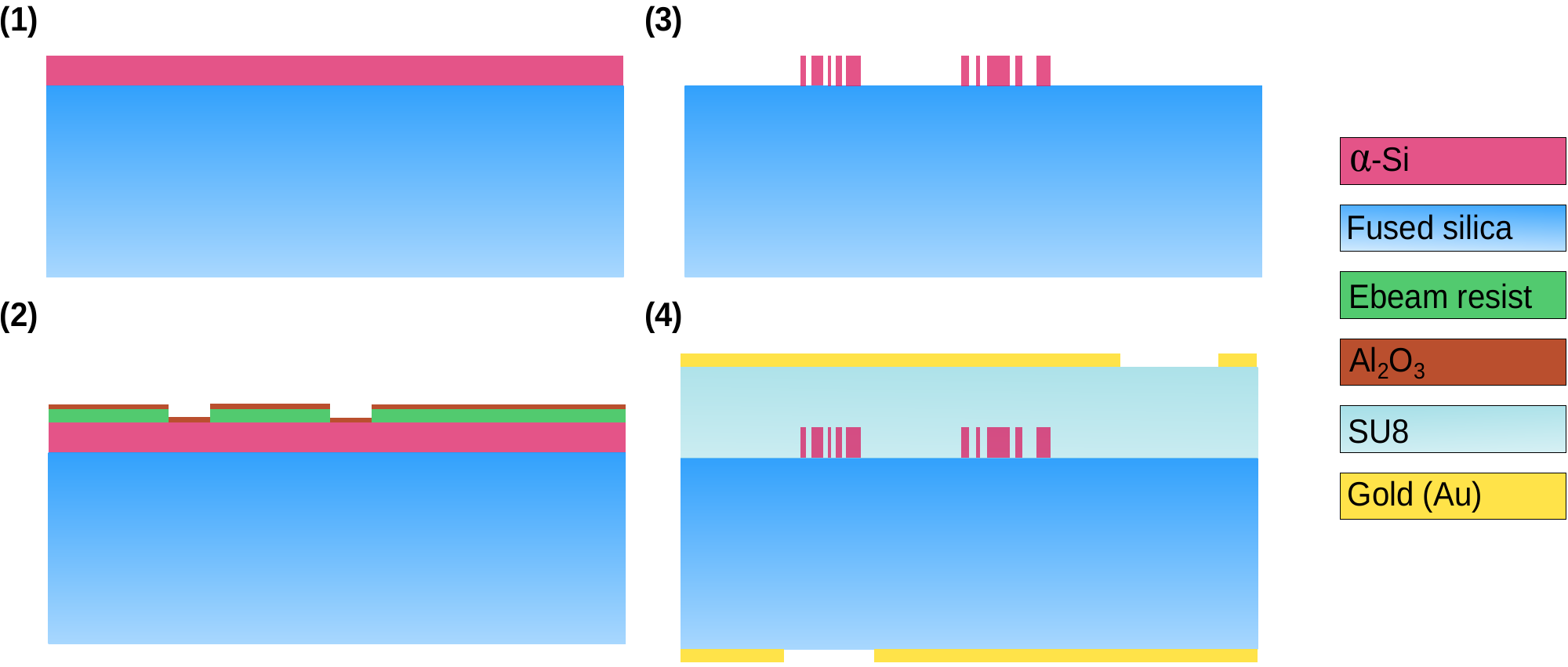}}
	\caption{\textbf{The key fabrication steps.} A 395-nm-thick layer of $\alpha$-Si is deposited on a 1-mm-thick fused silica substrate using PECVD. The metasurface pattern is generated with electron beam lithography, negated and transferred to the $\alpha$-Si layer via lift-off and dry-etching processes. Both sides are covered with and SU-8 layer, and the input and output apertures are defined through photolithography and lift-off. }
	\label{fig:S3_Fab}
\end{figure*}

\begin{figure*}[htbp]
	\centering
	\fbox{\includegraphics[width=\linewidth]{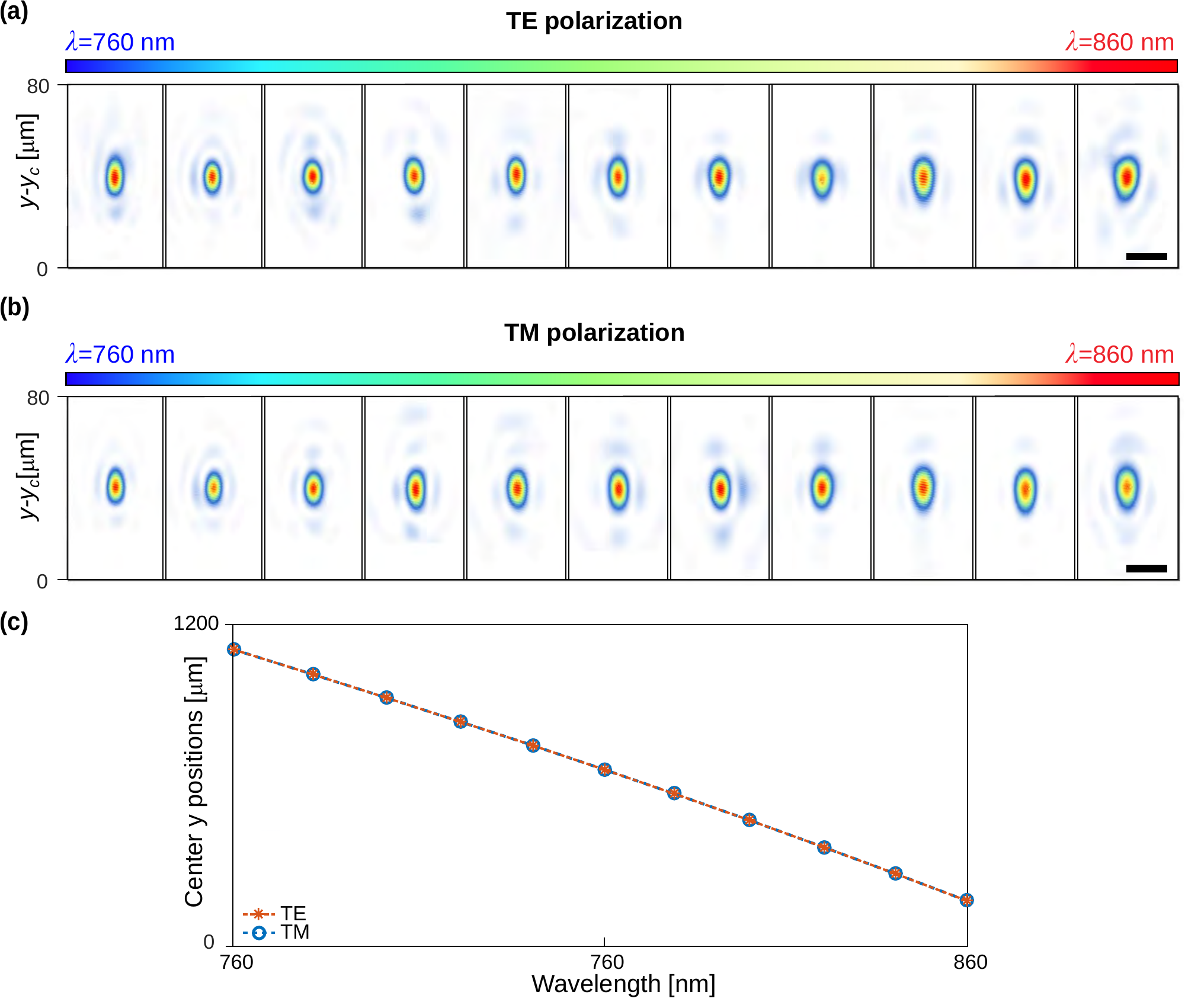}}
	\caption{\textbf{Focal plane intensity profiles.} (\textbf{a}) Two-dimensional intensity profiles measured at several wavelengths ($y_c$ is the center position of each profile) under illumination with TE polarized light, and (\textbf{b}) TM polarized light. (\textbf{c}) The position of the center of the focal spot along the dispersion direction, y, versus wavelength. The symbols represent the measured data, and the solid line is an eye guide. The scale bars are 20~$\mu$m.}
	\label{fig:S5_Focus}
\end{figure*}

\begin{figure*}[htbp]
	\centering
	\fbox{\includegraphics[width=\linewidth]{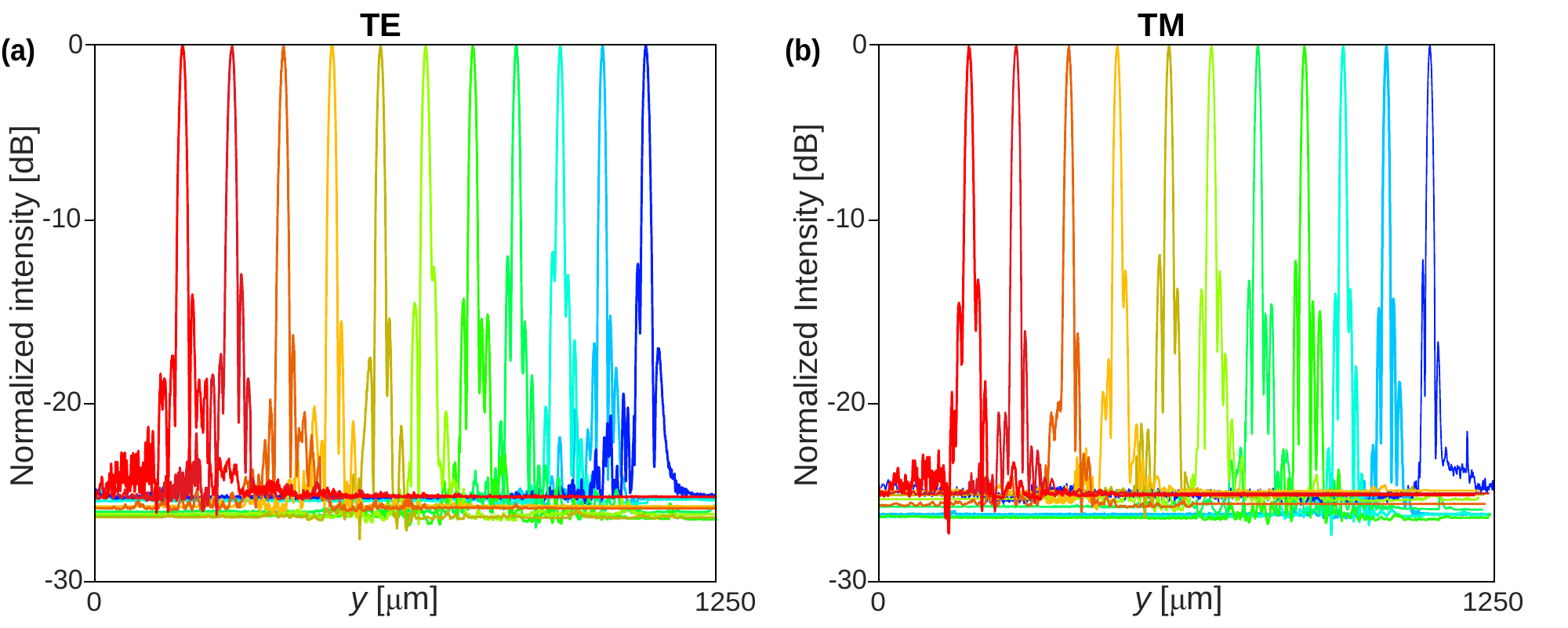}}
	\caption{\textbf{Intensity distribution profiles on logarithmic scales} (\textbf{a}) Same information as Fig. \textbf{4}b of the main text, plotted on logarithmic scale for TE polarization, and (\textbf{b}) for TM polarization.}
	\label{fig:S6_FocusLog}
\end{figure*}

\begin{figure*}[htbp]
	\centering
	\fbox{\includegraphics[width=0.5\linewidth]{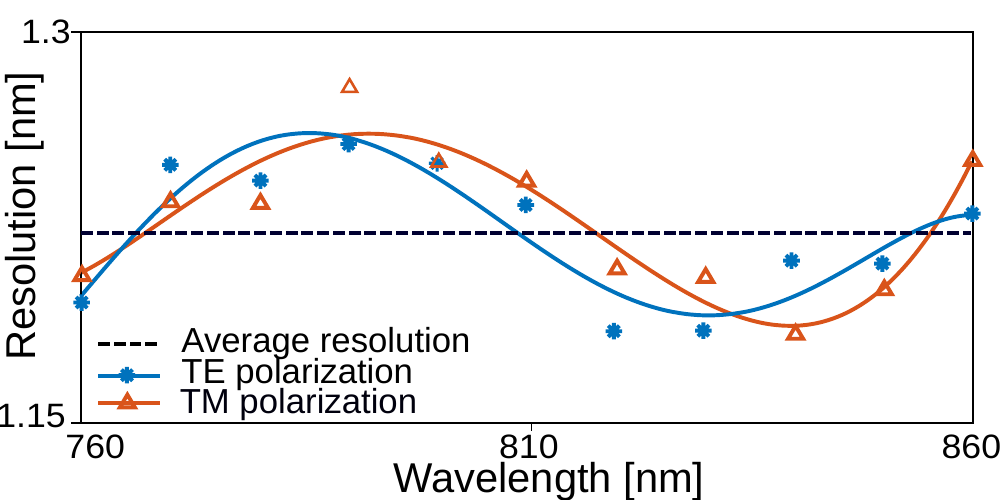}}
	\caption{\textbf{Measured spectral resolution versus wavelength.} Spectral resolution estimated using the measured focal spot FWHM and the displacement rate of the focal spot with changing the wavelength. The average resolution is 1.22~nm for both polarizations. The symbols show the measured date and the solid lines are eye guides.}
	\label{fig:S7_Res}
\end{figure*}

\textbf{This document contains supplementary information for "Compact Folded Metasurface Spectrometer''. Section S1 and S2 provides further details of the theoretical and experimental methods and apparatus used, and Figs. \ref{fig:S1_designH} to \ref{fig:S10_Res} contain further simulation and measurement data.} 

\section*{S1. Materials and Methods}
\label{sec:S1}
\subsection{Simulation and design} Ray tracing simulations of the spectrometer were performed using Zemax OpticStudio. In the simulations, metasurfaces were assumed to be phase-only diffractive surfaces. The grating was modeled as a blazed grating with a linear phase along the direction of dispersion (y), and independent of the other direction (x). The phase was chosen to correspond to a period of 1 $\mu$m, resulting in deflection angles of 31.6$^\circ$ and 36.35$^\circ$ at 760 nm and 860 nm, respectively. The angles were chosen such that the focused light could be captured by an objective with a numerical aperture of 0.95, while maximizing the dispersive power. The second and third surfaces were modeled as a summation of Cartesian coordinate polynomials (Binary~1), $\Sigma_{n,m} a_{m,n} x^my^n$, and cylindrical coordinate radially symmetric polynomials (Binary~2) $\Sigma_{i} b_{2i}\rho ^{2i}$. The coefficients were optimized to reduce geometric aberrations by minimizing the root mean square geometric spot radii for several input wavelengths covering the bandwidth. The optimized coefficients are given in Table \ref{tab:phase_profile}. As shown in to Fig. \textbf{2}b, all focal spots are optimized and are within the airy disks. This indicates that the designed spectrometer has small geometrical aberrations. The diffraction-limited resolution curve obtained is shown in Fig. \textbf{2}c. The simulations and optimizations were first performed in an unfolded configuration for simplicity. There were several constraints in finding the sizes for input and output apertures. Two opposing factors existed in determination of the 790-$\mu m$ input aperture diameter. On one hand, a larger input aperture results in a higher throughput and more captured light as well as a higher numerical aperture and potentially better resolution. On the other hand, the aperture size for the folded platform cannot be arbitrarily large because different metasurfaces should not overlap. Thus, the 790-$\mu m$ aperture diameter was chosen in the ray-tracing simulations as the largest size for which metasurface overlap can be avoided and diffraction-limited focusing can be achieved. The output aperture spatially filters the out of band wavelengths while passing through the bandwidth of interest. Therefore, its size was chosen as the smallest possible aperture that allows for all wavelengths of interest to pass through. Using the ray-tracing simulations, this optimum size was found to be 978 $\mu m$.
     
\begin{figure*}[htbp]
	\centering
	\fbox{\includegraphics[width=\linewidth]{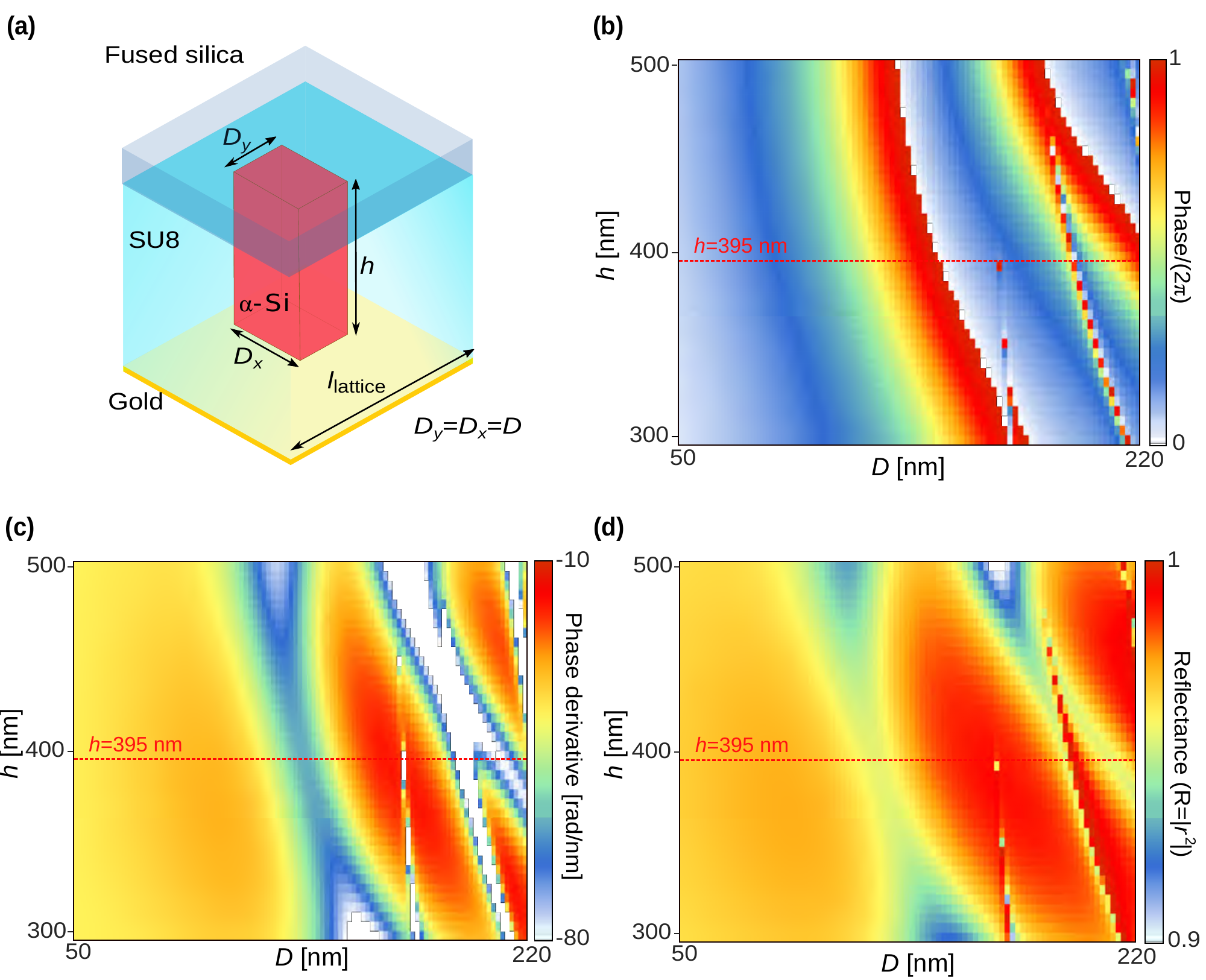}}
	\caption{\textbf{Single-post periodic lattice simulations.} (\textbf{a}) Schematic of a rectangular post on top of a fused silica substrate, showing the post dimensions. The nano-posts are capped by a 2-$\mu$m-thick layer of SU-8, and backed by a reflective gold layer. (\textbf{b}) Simulated reflection phase under TE illumination with $33.9^{\circ}$  incident angle. (\textbf{c}) Derivative of the phase with respect to the wavelength calculated and plotted versus the height ($h$) and width of the post ($D_x$$=$$D_y$$=$$D$). The nano-post height that provides full $2\pi$ phase coverage with high reflectance while minimizing variation of the phase derivative is found to be $h$=395~nm (the red line). (\textbf{d}) Reflectance as a function of post-width and height.}
	\label{fig:S1_designH}
\end{figure*}

The rigorous coupled wave analysis (RCWA) technique ~\cite{liu2012s} was used to obtain reflection phases of the nano-posts. For each specific set of dimensions, a uniform array of the $\alpha$-Si nano-posts was illuminated with a plane wave at the wavelength of 810 nm under an illumination angle of 33.9$^\circ$ and the reflected amplitudes and phases were extracted for each polarization. To choose the height of the nano-posts, we performed these simulations for nano-posts with square cross-sections and different heights and side lengths [Fig. \ref{fig:S1_designH}]. The height was then chosen to minimize the variation of the derivative of the phase with respect to wavelength for different side lengths, while providing a full 2$\pi$ phase coverage and high reflectivity. Considering the results of Figs. \ref{fig:S1_designH}b and \ref{fig:S1_designH}d, we chose the thickness to be 395 nm. Although this height is slightly less than $\lambda$/2, it is large enough to provide a full 2$\pi$ phase coverage as the device operates in reflection mode. The lattice constant was chosen to be 246 nm in order to satisfy the sub-wavelength condition and avoid higher order diffractions, which require $l_c$ $<$ $\lambda/n(1+\mathrm{sin(}\theta\mathrm{_{max})})$, where $l_c$ is the lattice constant, $n$ is the refractive index of the substrate, and $\theta_\mathrm{max}$ is the maximum deflection angle~\cite{kamali2016highly}. We chose $\mathrm{sin(}\theta_\mathrm{max}\mathrm{)}=1/n$, since light traveling at larger angles will undergo total internal reflection at the output aperture. To make the two focusing metasurfaces polarization-insensitive, reflection phase and amplitudes were obtained for nano-posts with rectangular cross section under oblique illumination with both TE and TM polarizations [Fig. \textbf{3}]. The design curves were then generated by determining a path in the $D_\mathrm{x}-$$D_\mathrm{y}$ plane along which TE and TM reflection phases are almost equal.

\begin{figure*}[htbp]
	\centering
	\fbox{\includegraphics[width=\linewidth]{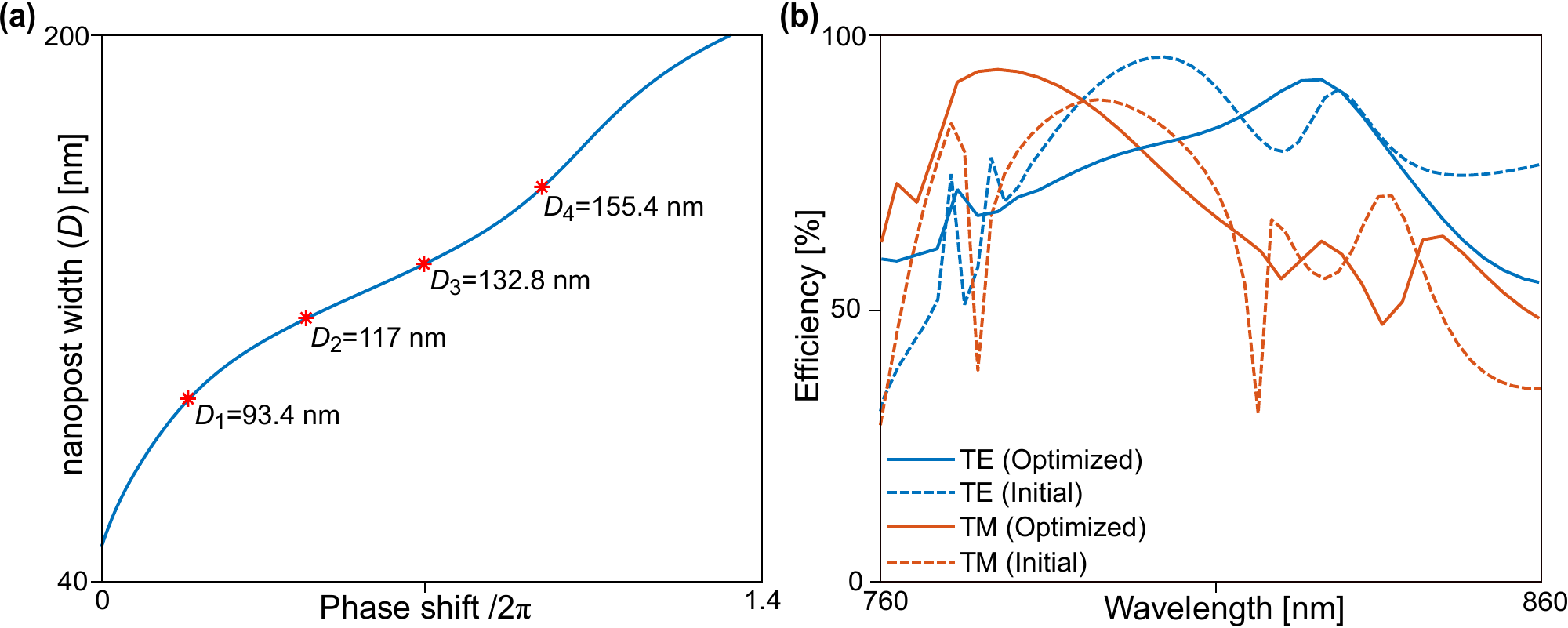}}
	\caption{\textbf{Grating design curves and deflection efficiencies.} (\textbf{a}) Post width versus reflection phase for 395-nm-tall posts on a square lattice with a 250-nm lattice constant. The red stars correspond to the nano-post sizes found from this graph that have the highest deflection efficiency over the bandwidth. (\textbf{b}) TE and TM polarization deflection efficiency curves versus wavelengths for the initial (i.e., the red stars in (a) ) and optimized nano-post dimensions given in Table \ref{tab:phase_profile1}.}
	\label{fig:S2_Grating}
\end{figure*}

For designing the blazed diffraction grating, we chose to use the same $\alpha$-Si thickness of 395 nm (for ease of fabrication). The lattice constant was set to be 250 nm, such that a grating period contains four nano-posts, and the structure becomes fully periodic. This allows for using periodic boundary conditions in the full-wave simulations of the structure, reducing the simulation domain size significantly. The initial values of the post widths were chosen using a recently developed high-NA metasurface design approach~\cite{arbabi2017increasing}. The simulation results for nano-post-width vs reflection-phase and the initial post widths are plotted in Fig. \ref{fig:S2_Grating}a. These values were then fed to a particle swarm optimization algorithm (using an RCWA forward solver) as a starting point. The algorithm optimizes the deflection efficiency of the grating for both polarizations at 11 wavelengths spanning the bandwidth of interest. The optimization parameters are the side lengths of the rectangular nano-posts, while their thickness and spacing is fixed. Deflection efficiencies of the initial and optimized gratings are plotted in Fig. \ref{fig:S2_Grating}b. The corresponding nano-post widths for both gratings are given in Table~\ref{tab:phase_profile1}.

\subsection{Sample fabrication} A summary of the key steps of the fabrication process is shown in Fig. \ref{fig:S3_Fab}. A 395-nm-thick layer of $\alpha$-Si was deposited on one side of a 1-mm-thick fused silica substrate through a plasma enhanced chemical vapor deposition process at 200$^\circ$C. The metasurface pattern was then generated in a $\sim$300-nm-thick layer of ZEP-520A positive electron resist (spun for 1 minute at 5000 rpm) using an EBPG5200 electron beam lithography system.  After development of the resist in a developer (ZED-N50, Zeon Chemicals), a $\sim$70-nm-thick alumina layer was evaporated on the sample in an electron beam evaporator. After lift-off, this layer was used as a hard mask for dry etching the $\alpha$-Si layer in a mixture of SF$_6$ and C$_4$F$_8$ plasma. The alumina layer was then removed in a 1:1 solution of  H$_2$O$_2$ and NH$_4$OH. A $\sim$2-$\mu$m-thick layer of SU-8 2002 polymer was spin-coated, hard-baked and cured on the sample to protect the metasurfaces. The output aperture (which is on the same side as the metasurfaces) was defined using photolithography (AZ-5214E positive resist, MicroChemicals) and lift-off. A $\sim$100-nm-thick gold layer was deposited as the reflective surface. To protect the gold reflector, a second layer of SU-8 2002 was used. To define the input aperture, a $\sim$2-$\mu$m-thick layer of SU-8 2002 polymer was spin-coated and cured on the second side of the wafer to improve adhesion with gold. The input aperture was then defined in a process similar to the output aperture.

\subsection{Device characterization procedure} 
\begin{figure*}[htbp]
	\centering
	\fbox{\includegraphics[width=\linewidth]{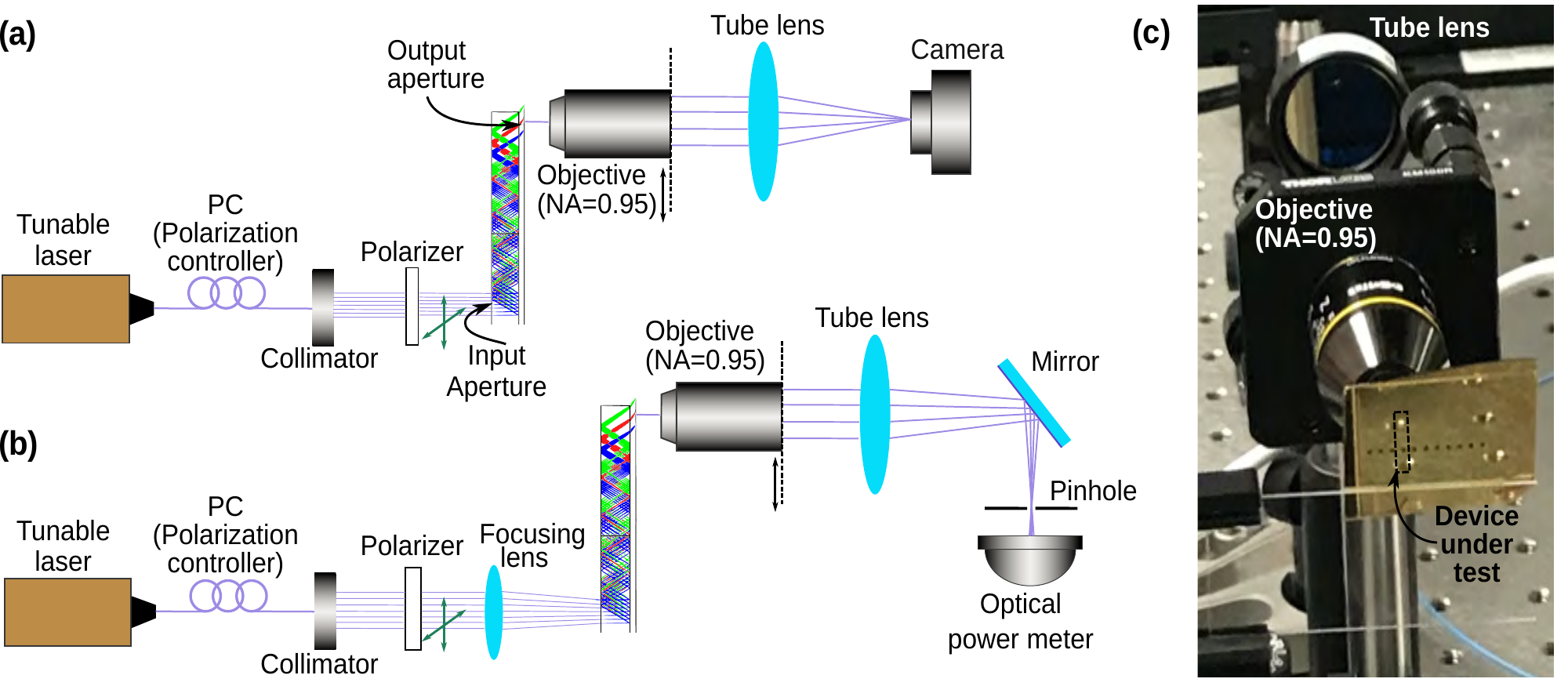}}
	\caption{\textbf{Measurement setups.}(\textbf{a}) Schematics of the measurement setup used for device characterization. (\textbf{b}) Schematics of the setup used to measure the focusing efficiencies. (\textbf{c}) An optical image of a part of the actual measurement setup showing the device, the objective lens, and the tube lens.}
	\label{fig:S4_setup}
\end{figure*}

The measurement setups used to characterize the spectrometer are schematically shown in Fig. \ref{fig:S4_setup}. Light from a tunable Ti-sapphire laser (SolsTiS, M-Squared) was coupled to a single mode optical fiber and collimated using a fiber collimator (F240FC-B, Thorlabs). A fiber polarization controller and a free space polarizer (LPVIS100-MP2, Thorlabs) were used to control the input light polarization, and different neutral density filters were used to control the light intensity. The beam illuminated the input aperture of the spectrometer at normal incidence. The focal plane of the spectrometer, located $\sim$200~$\mu$m away from the output aperture, was then imaged using a custom built microscope (objective: 100$\times$ UMPlanFl, NA=0.95, Olympus; tube lens: AC254-200-C-ML, Thorlabs; camera: CoolSNAP K4, Photometrics). Since the field of view is $\sim$136~$\mu$m (limited by the $\sim$15-mm image sensor, and the $\sim$111$\times$ magnification), while the total length over which the wavelengths are dispersed in the focal plane exceeds 1~mm, the objective is scanned along the dispersion direction to cover the whole focal plane at each wavelength (11 images captured for each wavelength). These images were then combined to form the full intensity distribution at each wavelength. The measurements were performed at 11 wavelengths (760~nm to 860~nm, 10-nm steps) to form the results shown in Figs. \textbf{4}b, \ref{fig:S5_Focus}, and \ref{fig:S6_FocusLog}. The measurements were also performed at a second set of wavelengths (761.25~nm, 811.25 nm, and 861.25~nm). These results are summarized in Fig. \textbf{4}d. The resolution [Fig. \ref {fig:S7_Res}] was estimated by finding the full-width-half-maximum (FWHM) at each wavelength, and  the displacement rate of the focus center along the y direction by changing the wavelength.
The setup was slightly changed for measuring the focusing efficiencies. The input beam was partially focused by a lens ($f=$10~cm) such that all the beam power passed through the input aperture (with a $\sim400$ $\mu$m FWHM). In addition, the camera was replaced by a photodetector and a pinhole with a diameter of 3.5~mm in front of it to measure the focused power. The pinhole, corresponding to a $\sim$31-$\mu$m area in the focal plane, allows only for the in-focus light to contribute to the efficiency. The efficiency is then calculated at each wavelength by dividing these measured powers by the total power tightly focused by a 10-cm focal length lens that was imaged onto the power meter using the same microscope (i.e., by removing the spectrometer and the pinhole). 
The experimental setup for capturing the sample spectra is almost identical to Fig. \ref{fig:S4_setup}b, with the only difference of the polarizer being replaced by the sample of interest, and an 840-nm short-pass filter inserted before the sample. The light source was also replaced by a supercontinuum laser (Fianium Whitelase Micro, NKT Photonics).

\subsection{Angular response measurement} 
\begin{figure*}[htbp]
	\centering
	\fbox{\includegraphics[width=\linewidth]{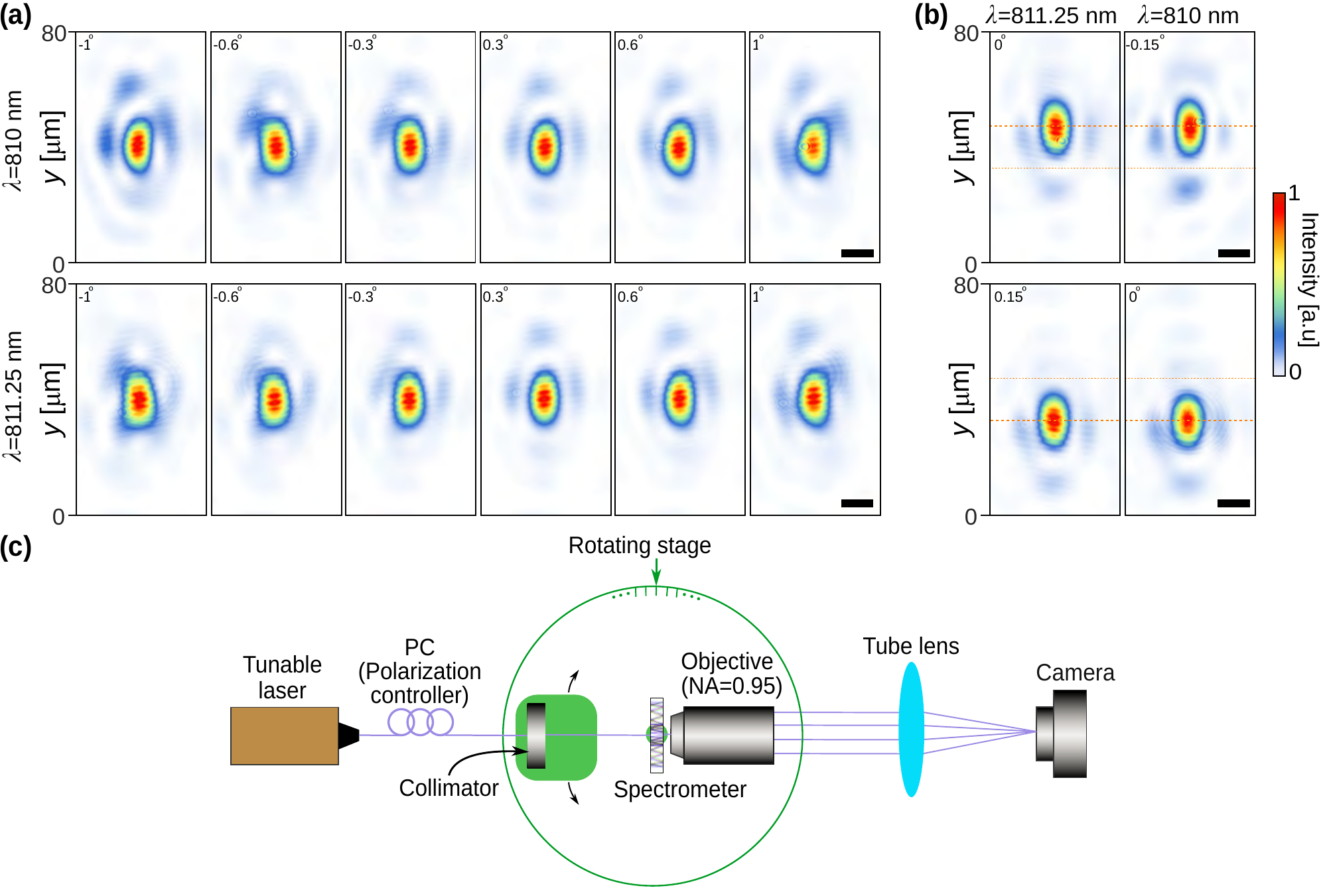}}
	\caption{\textbf{Measured angular response of the device for polar angle variation with respect to 0 angle in $x-z$ and $y-z$ planes.}  (\textbf{a}) Angular response of the device for different tilted incident angles between -1$^\circ$ to +1$^\circ$ in the $x-z$ plane. (\textbf{b}) Angular response of the device  for $\pm$0.15$^\circ$ tilted incident angles in the $y-z$ plane. (\textbf{c}) experimental setup used for measuring the angular response. The scale bars are 10~$\mu$m.}
	\label{fig:S9_Ang}
\end{figure*}

To measure the angular response of the device we used the setup shown in Fig. \ref{fig:S9_Ang}c, equipped with a rotating stage with 0.1$^\circ$ precision in the $x$-$z$ plane and 0.002$^\circ$ in the $y$-$z$ plane. The collimator (connected to the fiber coming from the source) was mounted on this rotating stage, where the folded spectrometer was exactly located at its center. The incident angles were adjusted accordingly for 0$^\circ$, $\pm$0.3$^\circ$, $\pm$0.6$^\circ$, $\pm$1$^\circ$ angles. As can be observed in Fig. \ref{fig:S9_Ang}a, the focal spots did not vary much in size as the angle is sweeped between -1$^\circ$ to +1$^\circ$ in x-direction. For measuring the tilt angle in the $y$-$z$ plane, the distance from the collimator to the device was measured to be 280 mm. In order to impose $\pm$0.15$^\circ$ tilt in y direction, the mounted collimator level is raised or lowered by 0.73 mm, and its tilt was adjusted accordingly such that the beam hits the center of the input aperture. As shown in Fig. \ref{fig:S9_Ang}b, such a tilt in input incident angle does not degrade the spectral resolution of 1.25 nm.
\section*{S2. Increased Throughput Design}
\begin{figure*}[htbp]
	\centering
	\fbox{\includegraphics[width=\linewidth]{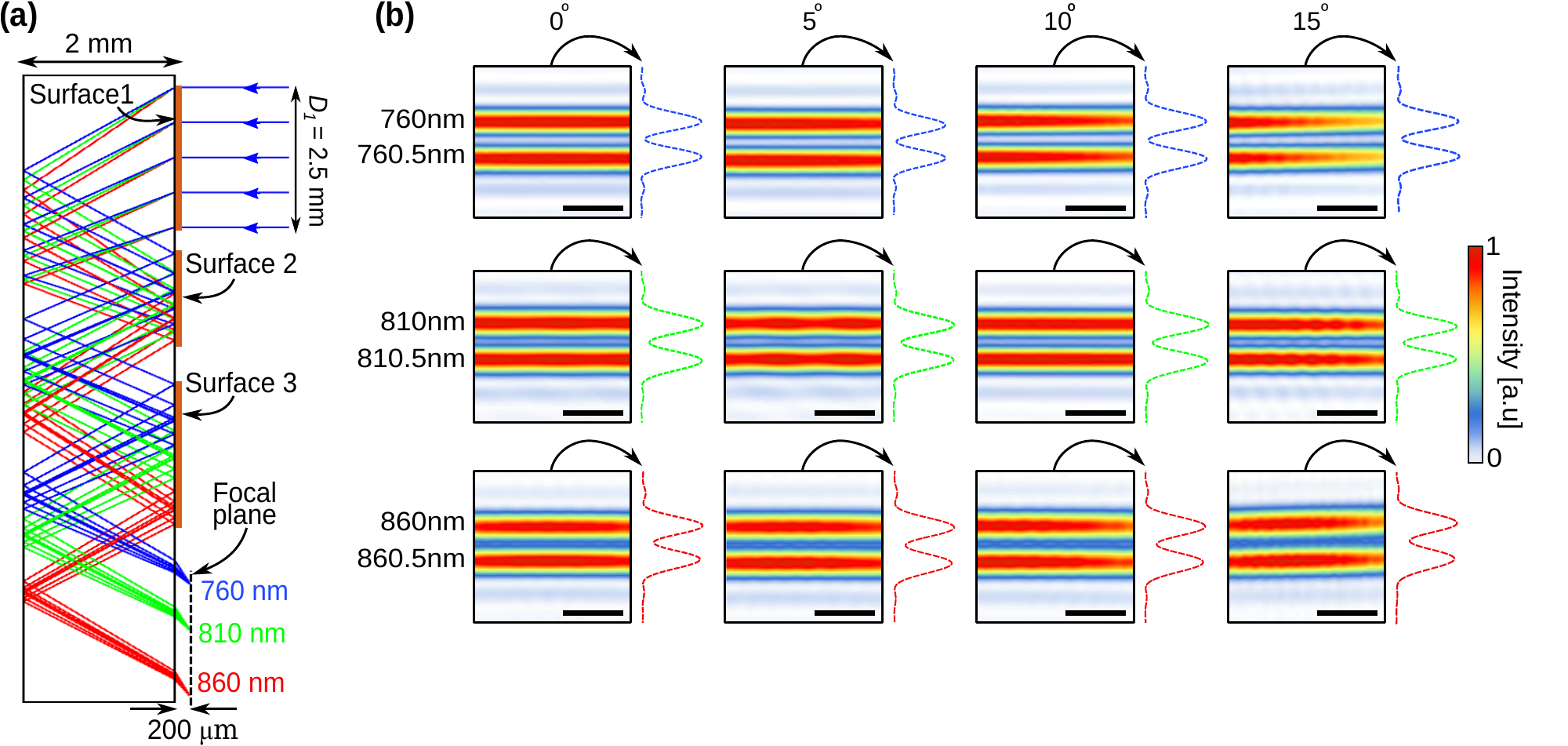}}
	\caption{\textbf{Ray-optics design and simulation results of an extended-throughput folded spectrometer.} (\textbf{a}) Ray tracing simulation results of the extended-throughput folded spectrometer, shown at three wavelengths in the center and two ends of the band. The system consists of three metasurfaces optimized to separate different wavelengths of the light and focus them on the focal plane. (\textbf{b}) Simulated intensity distribution for two wavelengths separated by 0.5 nm around three different center wavelengths of 760 nm, 810 nm, and 860 nm for 4 different incident angles of 0$^\circ$, 5$^\circ$, 10$^\circ$ and 15$^\circ$. The intensity distributions show that wavelengths separated by 0.5 nm are theoretically resolvable for all aforementioned incident angles. The scale bars are 15~$\mu$m.}
	\label{fig:S10_Res}
\end{figure*}

To further demonstrate the capabilities of the platform, we have designed a second spectrometer with significantly increased throughput. In order to achieve higher throughput, a larger input aperture was required, so the slab thickness was increased to 2 mm to give more freedom on the non-overlapping condition for the metasurfaces. The design, as shown in Fig. \ref{fig:S10_Res}, has a 2.5mm input aperture. To further improve the throughput, the acceptance angle of the device was increased. To achieve this goal, we took an approach similar to  the fabricated spectrometer with the difference of adding extra phase terms to the input diffraction grating. This helps with orienting the focuses on the image plane for different incident angles, as well as relaxing the condition for focusing in the x-direction. This in turn allows for increasing the input incident angle to $\pm15^\circ$ degrees. The phase profile coefficients for metasurfaces 1 to 3 in Fig. \ref{fig:S10_Res} are given in Table \ref{tab:phase_profile_extended}. In the final design, the power is distributed in an area close to 200 $\mu$m wide in the x-direction in the focal plane, instead of a diffraction limited focus. According to the intensity profiles shown in Fig. \ref{fig:S10_Res}b, the device can distinguish between wavelengths spaced by 0.5 nm both at the center wavelength of 810nm, and also at the side wavelengths of 760 nm and 860 nm. Based on the angular response of the device in the x-z and y-z planes, and also the input aperture size of the device, an etendue of around ~$\sim$13000 $Sr(\mu m)^2$ is estimated.

\bibliographystyle{naturemag}
\bibliography{References}

\pagebreak

\begin{table*}[htbp]
	\centering
	\caption{\bf Phase profile coefficients in terms of $[rad/mm^{m+n}]$ for metasurfaces 1 and 2}
	
\scalebox{0.98}{
\begin{tabular}{|ccccccccccc|}
	\hline
	Metasurface   & $a_{x^2y^0}$ & $a_{x^0y^2}$ & $a_{x^2y^1}$ & $a_{x^0y^3}$ & $a_{x^4y^0}$  & $a_{x^2y^2}$  & $a_{x^0y^4}$ &  $a_{\rho^6}$  & $a_{\rho^8}$   & $a_{\rho^{10}}$ \\
	\hline
	I (R=525.0 $\mu$m)  & $-4.02$ & $-2.08$ & $0.47$ & $0.20$ & $-5.68e-4$ & $7.55e-3$ & $2.36e-3$  & $1.93e-4$ & $-3.22e-6$  & $-5.886e-9$ \\
	II (R=582.5 $\mu$m) & $-3.91$ & $-3.70$ & $-0.68$ & $-0.24$ & $ 6.26e-3$ & $0.021$ & $6.34e-3$  & $-2.48e-4$ & $4.82e-6$  & $-2.974e-9$ \\
	\hline
\end{tabular}}
	\label{tab:phase_profile}
\end{table*}

\begin{table*}[htbp]
	\centering
	\caption{\bf Optimized grating post sizes [nm] ($D_x, D_y$)}
	\begin{tabular}{|c|cc|cc|cc|cc|}
		\hline
		Optimization  & $D_{x1} $ & $D_{y1}$ & $D_{x2}$ & $D_{y2}$ & $D_{x3}$ & $D_{y3}$ & $D_{x4}$ & $W_{y4}$\\
		\hline
		Maximizing first order diffraction Efficinecy & $93.4$ & $93.4$ & $117$ & $117$ & $132.8$ & $132.8$ & $155.4$ & $155.4$   \\
		Particle Swarm Optimization & $68$ & $134$ & $115.2$ & $119.6$ & $147.4$ & $151.2$ & $137.8$ & $178.8$ \\
		\hline
	\end{tabular}
	\label{tab:phase_profile1}
\end{table*}

\pagebreak

\begin{table*}[hb]
	\centering
	\caption{\bf Phase profile coefficients of the increased throughput design (Fig. \ref{fig:S10_Res}) in terms of $[rad/mm^{m+n}]$}
	
	\scalebox{0.98}{
		\begin{tabular}{|ccccccccccc|}
			\hline
			Metasurface   & $a_{x^2y^0}$ & $a_{x^0y^2}$ & $a_{x^2y^1}$ & $a_{x^0y^3}$ & $a_{x^4y^0}$  & $a_{x^2y^2}$  & $a_{x^0y^4}$ &  $a_{\rho^6}$  & $a_{\rho^8}$   & $a_{\rho^{10}}$ \\
			\hline
			I (R=2.5 mm)  & $-0.44$ & $-2.98$ & $0.015$ & $0.035$ & $-1.5$e$-5$ & $-6.16$e$-5$ & $-3.1$e$-4$  & $1.42$e$-7$ & $-2.26$e$-10$  & $-5.886$e$-13$ \\
			II (R=1.5 mm) & $-3.13$ & $1.91$ & $0.027$ & $-0.04$ & $-8.2$e$-4$ & $-0.72$e$-4$ & $3.8$e$-3$  & $1.12$e$-5$ & $-5.26$e$-8$  & $7.58$e$-11$ \\
			III (R=2.15 mm) & $-2.45$ & $-2.69$ & $-0.067$ & $-0.06$ & $3.56$e$-4$ & $-1.73$e$-5$ & $-9.14$e$-4$  & $6.71$e$-7$ & $-1.72$e$-9$  & $1.62$e$-12$ \\
			\hline
	\end{tabular}}
	\label{tab:phase_profile_extended}
\end{table*}